# 2D Single Crystal of High-Temperature Phase Cuprous Iodide under Ambient Conditions

Bingquan Peng[1*], Jie jiang[2], Fangfang Dai[1], Liang Chen[2], and Lei Zhang[3]

[1]Wenzhou Institute, University of Chinese Academy of Sciences, Wenzhou, Zhejiang 325000, China

[2]School of Physical Science and Technology, Ningbo University, Ningbo, 315211, China

[3]MOE Key Laboratory for Nonequilibrium Synthesis and Modulation of Condensed Matter, School of Science, Xi'an Jiaotong University, Xi'an 710049, China

*Corresponding authors. Email: pengbq@ucas.ac.cn;

**Abstract**

**Two-dimensional (2D) materials, with their structural uniqueness, exceptional properties, and wide-ranging applications, show unprecedented prospects in fundamental physics research and industrial applications. 2D β-phase cuprous iodide (β-CuI) is a promising candidate for overcoming the challenges of insufficient P-type transparent conductive materials, with multiple predicted unique properties. Previous experimental results show that β-CuI only occurs at elevated temperatures between 645 and 675 K. Many efforts are made to stabilize β-CuI at room temperature through surface/interface engineering. However, the resulting composites limit the performance and application of pure 2D β-CuI. Here, we demonstrate experimentally that isolated 2D β-CuI single crystals can exist stably under ambient conditions, a high-temperature phase CuI found at room temperature. We validate the simultaneous existence of γ-CuI and β-CuI in the synthesized CuI. The previous neglect of β-CuI crystals can be ascribed to factors including their low content, small dimensions, and lack of ingenious experimental characterization. Moreover, the theoretical calculation further confirms dynamically and thermally stable of the monolayer β-CuI, which has an**

**ultra-wide direct band-gap (3.66 eV). Our findings challenge the traditional understanding of β-CuI as a high-temperature phase of CuI, instead providing a new definition that 2D β-CuI exhibits remarkable stability under ambient conditions.**

**Keywords**: 2D single crystal, layered β-CuI, high-temperature phase, P-type transparent conductive materials, 2D ultrawide bandgap semiconductors

Copper iodide (CuI) has garnered much attention as a highly attractive p-type semiconductor with wide-bandgap and high conductivity, making it an essential component in various fields such as transparent flexible electrodes, thermoelectric devices, p-channel transistors, light-emitting diodes, catalysts, and solar cells [1-7]. CuI has a complex phase diagram, exhibiting three stable crystal forms, namely, α-, β-, and γ-CuI [8-10]. The stable form of CuI is believed to be the zinc blende structure (γ-CuI) below 643 K [3], while the cubic α-CuI structure dominates above 713 K [9,11-12]. The layered β-CuI with a wurtzite structure is only stable within a narrow temperature range of 645-675 K, also called the high-temperature phase CuI. Considering that 2D materials usually exhibit superior electrical, optical and other physical properties than bulk materials [13-17], 2D CuI is still attracting a lot of theoretical research, although it is considered as a high-temperature phase layered material. Theoretical research suggests that 2D β-CuI exhibits unique properties, including low thermal conductivity and an ultra-wide direct bandgap, as compared to the bulk γ-phase cuprous iodide [18]. With its rough surface, γ-CuI limits the practical application of the material in transparent electronics, whereas β-CuI possesses promising potential as a transparent electronic material in 2D applications [18-20]. Additionally, first-principles calculations predict a structure of β-phase CuI as a topological unconventional Dirac semimetal, further highlighting its potential for novel applications [21]. Unfortunately, none of these unique properties of β-CuI are experimentally verified, mainly because it is not synthesized or found at room temperature.

The existence of layered β-CuI is confined to a relatively narrow high-temperature range, thereby impeding the progress of experimental investigation and the development of applications for its unique properties. Therefore, this also inspires the development of methods to stabilize β-CuI at room temperature. These methodologies are mainly based on surface/interface engineering, including graphene encapsulation, synthesis on atomically smooth graphite basal plane surfaces using a hybrid electrochemical/chemical method, and deposition on specific substrates via close distance sublimation [10,22-24]. It should be noted, however, that the stabilized β-CuI produced using these methods is a composite material. Although novel properties such as the piezoresistive effect and room-temperature ferromagnetism can be observed in reduced graphene oxide films encapsulating 2D β-CuI crystals, these findings do not fully reflect the intrinsic characteristics of pure 2D β-CuI crystals [24]. In recent years, theoretical predictions were made about the stability of 2D β-CuI, suggesting that it can be stable at room temperature, which brings hope and motivation to find or synthesize it experimentally [18,25]. However, to date, the synthesis or discovery of its single crystal structure under ambient conditions remains elusive.

In this study, we report the experimental discovery of isolated single crystals of 2D β-CuI on a micro-nano scale under ambient conditions. These typical hexagonal crystals were confirmed to be β-CuI by high-resolution transmission electron microscopy (HRTEM), TEM-energy-dispersive X-ray spectroscopy (TEM-EDS), selected area diffraction (SAED), and diffraction pattern simulation and analysis. The atomic-resolution imaging of the crystal structure of 2D β-CuI was obtained directly by HRTEM. Our experiments revealed the presence of a Moiré superlattice at the overlap region of the samples, indicating that the observed CuI is a layered crystal with atomic thickness, i.e. 2D β-CuI. Moreover, the theoretical simulation of this superlattice using the Angle theory of phase diagram was consistent with our experimental results. Ab initio molecular dynamics (AIMD) simulations show that monolayer β-CuI is thermally stable at 300 K. Meanwhile, theoretical studies show that monolayer β-CuI is dynamically stable, with an ultra-wide direct band-gap (3.66 eV). According to the Material Project, the monolayer β-CuI crystal structure with P-3m1 symmetry, discovered in this work, is not currently identified as a material that matches experimentally determined crystal structures [26]. Additionally, the experimental results show that a very small amount of 2D β-CuI coexists with a large amount of γ-CuI under ambient conditions after CuI is synthesized. To the best of our knowledge, this discovery represents the initial experimental confirmation of the stable existence of high-temperature phase CuI single crystals under ambient conditions. This discovery will also redefine the phase diagram of CuI crystal structure at different temperatures, indicating that the high-temperature phase β-CuI can stably exist under ambient conditions.

2D β-CuI crystals were obtained by grinding and ultrasonic treatment of CuI powder, as depicted in **Figure 1**a. These typical hexagonal crystals tend to be randomly distributed on an ultra-thin carbon film coated Mo TEM grid, as shown in Figure 1c and Figure S2 (Supporting Information). Initially, a full elemental analysis of the hexagonal region was carried out through TEM-EDS. Element mapping revealed that iodine and copper were associated with the typical hexagonal region, indicating that the region mainly consisted of CuI crystals with a Cu:I ratio of approximately 1:1, as illustrated in Figure 1d and Figure S3. The contrast of the typical hexagonal region in the TEM image at low magnification was similar to that of the supported ultra-thin carbon film, suggesting that the sample thickness was very thin. A STEM image (HAADF) was able to clearly show the typical hexagonal shape of the sample, as demonstrated in Figure 1d.

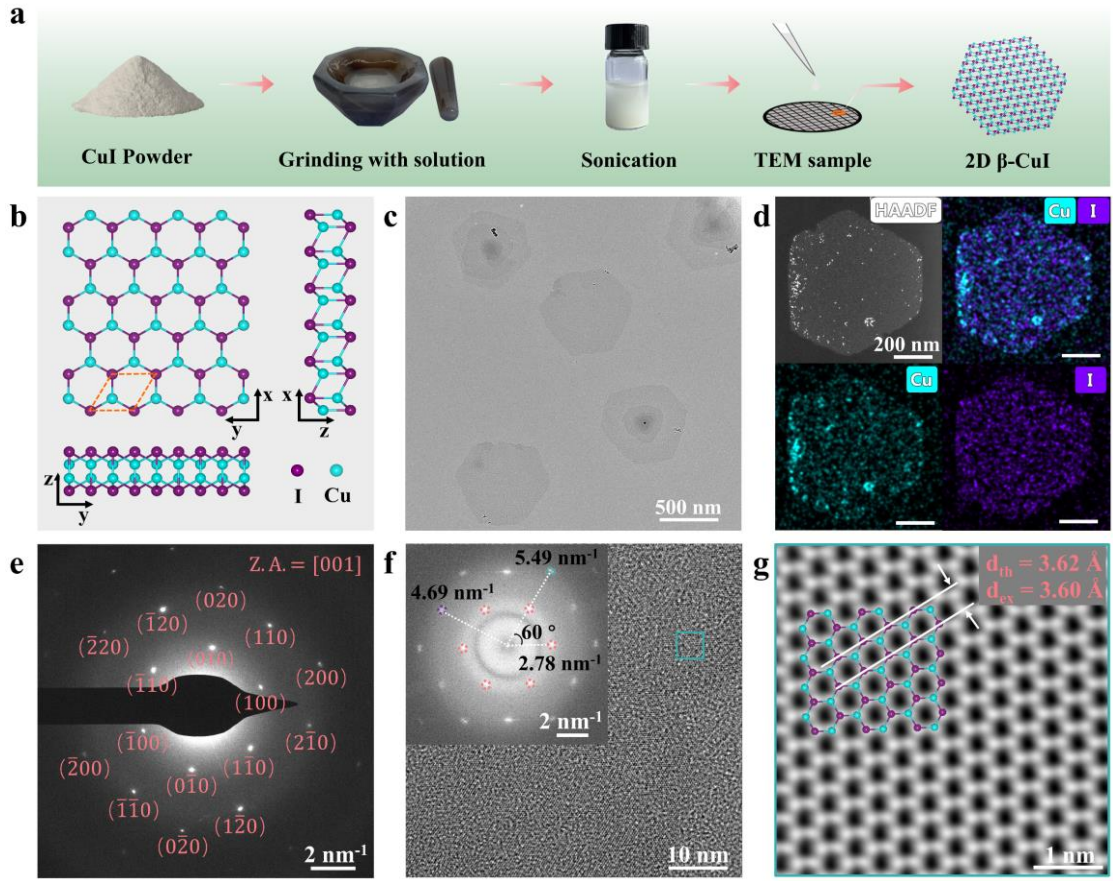

**Figure 1.** Schematic of the preparation process, atomic structure, morphology, element mapping, and characterization for 2D β-CuI single crystals. (a) Schematic drawings of the TEM sample preparation processes. (b) Atomic structure of 2D β-CuI with the primitive cell outlined. This structure exhibits P-3m1 symmetry (space group: 164), with a lattice constant of 4.18 Å. (c) Bright-field TEM image of 2D β-CuI single crystals on a TEM support film. (d) STEM image and Energy-dispersive X-ray spectroscopy (EDS) mapping of 2D β-CuI single crystal. (e) The selected area diffraction (SAED) pattern of one β-CuI crystal in (c), exhibits a hexagonal lattice pattern with six first-order maxima points at $2.79 \pm 0.01$ nm$^{-1}$, highlighting Bragg spots in the [001] zone axis of the β-CuI single crystals. (f) The high-resolution TEM of a β-CuI crystal. (e) The FFT of the entire bright-field image (f). (g) This magnified area is extracted from the cyan box in image (f), showing a 2D β-CuI crystal structure module where Cu and I atoms are represented as cyan spheres and purple spheres, respectively. Theoretical and experimental lattice spacings are indicated as $d_{th}$ and $d_{ex}$, respectively. The image

is obtained by applying a spatial filter to the original image (f).

The sets of six-fold symmetric SAED patterns for an individual hexagonal sample region indicate that the sample is single crystal, as shown in **Figure 1**e. The previous diffraction analysis of 2D β-CuI in rGO membrane led us to believe that the SAED displays a single β-CuI at [001] zone axis, with a hexagonal lattice pattern exhibiting six first-order maxima points corresponding to a lattice spacing of 3.59 ± 0.01 Å [24]. A lattice-resolved TEM image in Figure S1i confirms that the lattice with hexagonal symmetry (3.59 ± 0.07 Å) is consistent with the lattice obtained using SAED. Previous research suggests that CuI can only exist as a γ-phase at room temperature [22]. To further confirm the experimental results, we simulated the electron diffraction patterns of β-CuI and γ-CuI using CrysTBox software [27]. The diffraction pattern in Figure 1e is found to be a [001] zone-axis diffraction pattern for β-CuI (see Supplementary Information Section 3), thus confirming that the CuI crystals are β phase crystals in this work. Furthermore, we confirmed the constant of this lattice through fast Fourier transform (FFT) images from high-resolution TEM, as shown in Figure 1f. Performing a bandpass-filtered inverse FFT (iFFT) discarded lower frequencies and removed aperiodic pixel noise [28]. The atomic-resolution TEM images of the β-CuI crystal were obtained in Figure 1g, displaying a close correspondence with the atomic model of the 2D β-CuI crystal structure. It is worth noting that Cu and I atoms cannot be distinguished in the HRTEM image, and the model was solely used for matching the crystal structure. In this study, the stability of β-CuI under electron beam irradiation allowed us to successfully obtain atomically resolved images (see Movie S1 and Figure S5).

The moiré pattern is a unique feature of 2D layered materials that arises from the interference effect of atomic lattices, setting them apart from bulk materials [29-30]. In this study, TEM observations revealed the presence of moiré structures in 2D β-CuI crystals, as depicted in **Figure 2**a. Based on FFT analysis, it was found that two single crystals were stacked together in different orientations, with a relative rotation angle of ~16°. For homobilayers, the unit cell of such a moiré superlattice can be controlled by the twisting angle. Therefore, a moiré superlattice was created by twisted homobilayers, which was generated by modulating the twisting angle (θ =16°) on lattices of the same material (2D layered β-CuI). On the basis of theoretical model, the unit cell of the superlattice obtained by rotating the two layers of β-CuI with a rotation of 16° was consistent with the experimental results, as shown in Figure 2b and 2c. These results indicate that 2D β-CuI with atomic thickness was successfully obtained in this work.

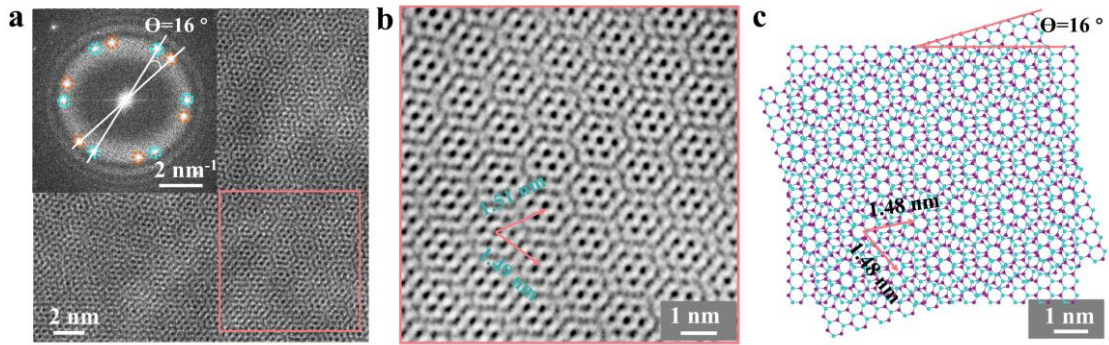

**Figure 2**. Moiré superlattice in 2D β-CuI crystals. (a) This image presents the moiré pattern caused by the twist angle between two layers of β-CuI. The rotation angle of two layers of β-CuI can be observed in the FFT pattern (inset), which is estimated as 16°. (b) iFFT of the area selected in pink in g, which is magnified in the pink box. (c) Illustration of β-CuI superlattice formation by moiré rotation. The observed periodicity can be explained by rotation of the topmost β-CuI layer.

Based on the P-3m1 symmetry of the monolayer β-CuI atomic structure shown in Figure 1b (space group :164), the lattice constant of 4.15 ± 0.01 Å can be obtained from lattice spacing data in TEM images (Figure 1f), which is almost consistent with the theoretical value of 4.18 Å. However, it should be noted that different studies have reported slightly different lattice constants for β-CuI, indicating that the lattice constant can be affected by environmental factors such as temperature or interface. For instance, Sakuma measured a lattice constant of 4.28 Å for β-CuI at 693 K using X-rays [11], while Keen and Hull reported a lattice constant of 4.30 Å for β-CuI at 655 K in neutron diffraction experiments [12]. Hsiao et al. reported a lattice constant of 4.27 Å for β-CuI on the atomically smooth graphite basal plane surface using TEM-SAED analysis [23]. More recently, Mustonen et al. measured a lattice constant of 4.19 ± 0.07 Å for a monolayer β-CuI encapsulated in graphene at room temperature using nano-beam electron diffraction [22]. Despite these variations, the lattice constant of 4.15 ± 0.01 Å observed in this study is still close to the reported values, indicating that environmental factors can indeed influence the lattice constant of β-CuI.

As revealed by X-ray diffraction (XRD) technique, the crystal phase transition of CuI with temperature can be summarized as follows: γ-CuI is stable below 643 K, β-CuI is stable between 645 and 675 K, and α-CuI is stable above 713 K [11]. A similar phase transition occurs in silver iodide (AgI), where γ-AgI remains stable up to 408 K, β-AgI is stable between 408 and 419 K, and α-AgI is stable above 419 K [31]. However, when AgI is synthesized directly in solution, it exists as a coexistence of both γ-AgI and β-AgI [32]. The synthesis of metastable phases, such as β-CuI and β-AgI, is influenced by various factors including pressure, temperature, and size. The identification and synthesis of the increasing number of possible or hypothesized metastable crystal materials is an immensely challenging task, primarily due to the lack of a rigorous metric or criterion for determining which compounds can be synthesized [33-34]. γ-CuI undergoes a phase transition to β-CuI during heating in the range of 645 K to 675 K [35]. However, this fact alone cannot determine whether β-CuI can exist stably under ambient conditions, as its stability is influenced by multiple factors such as temperature, pressure, chemical environment, etc. The search for metastable materials has predominantly been a heuristic process, relying on empirical observations, intuition, and at times speculative predictions, often referred to as "rules of thumb. [17,36]" Therefore, the discovery of stable β-CuI under ambient conditions has led to an updated understanding of the metastable structure of β-CuI.

Furthermore, in order to further elucidate the origin of β-CuI, we synthesized CuI

through the reaction between hydrogen iodide and cupric chloride, followed by a comprehensive crystallographic analysis (see Supplementary Information Section 4). Firstly, analysis of the synthesized white precipitate was performed using Powder XRD. In Figure S11, it was found that the XRD peaks of the sample matched those of the purchased standard CuI powder, confirming the synthesis of CuI. Upon comparison with Figure S1, it can be concluded that the main component of the sample is γ-CuI. Subsequently, irregular nano-sheets were identified in the TEM analysis, as shown in Figure S12a. Elemental analysis using EDS-TEM confirmed that these nano-sheets consist of CuI (see Figures S12 and S13). The diffraction patterns and high-resolution TEM analysis indicated that these nano-sheets correspond to β-CuI. Certainly, γ-CuI was also observed through TEM analysis, as depicted in Figure S14. These results suggest that the dominant phase in the synthesized CuI is γ-CuI, with some nano-sheets being β-CuI. Interestingly, conventional Powder XRD was not able to reliably detect β-CuI in the sample. These findings are consistent with the results obtained from commercially purchased standard CuI powder, providing further evidence for the presence of β-CuI under ambient conditions. From the results obtained, it can be concluded that the synthesis process of CuI involves the simultaneous presence of two structural forms, namely γ-CuI and β-CuI. The previous neglect of β-CuI crystals can be attributed to various factors, including their low content, small dimensions, and the lack of suitable sample preparation methods and reliable experimental characterization techniques.

   The unexpected discovery of such high-temperature phase crystals at room temperature prompted us to analyze and discuss the stability of 2D β-CuI through DFT theoretical calculation and AIMD simulation. Firstly, we confirmed the thermally stable of monolayer β-CuI by AIMD simulation. Through a long-term AIMD simulation performed at 300 K, we observed that the crystal maintained a relatively stable structure, as depicted in **Figure 3**a. The radial distribution functions (RDFs) between Cu and I exhibited obvious crystal peaks, and the probability density between the first two peaks was almost zero, indicating that the ionic bond between copper and iodine remained intact (Figure 3b). Additionally, the root-mean-square deviation (RMSD) values of Cu ion and I ion remained stable, as shown in Figure 3c. By analyzing the bond length and angle of the structure at different times, we found that their characteristics were consistent (Figures 3d and 3e). Therefore, we conclude that 2D β-CuI is thermodynamically stable at room temperature.

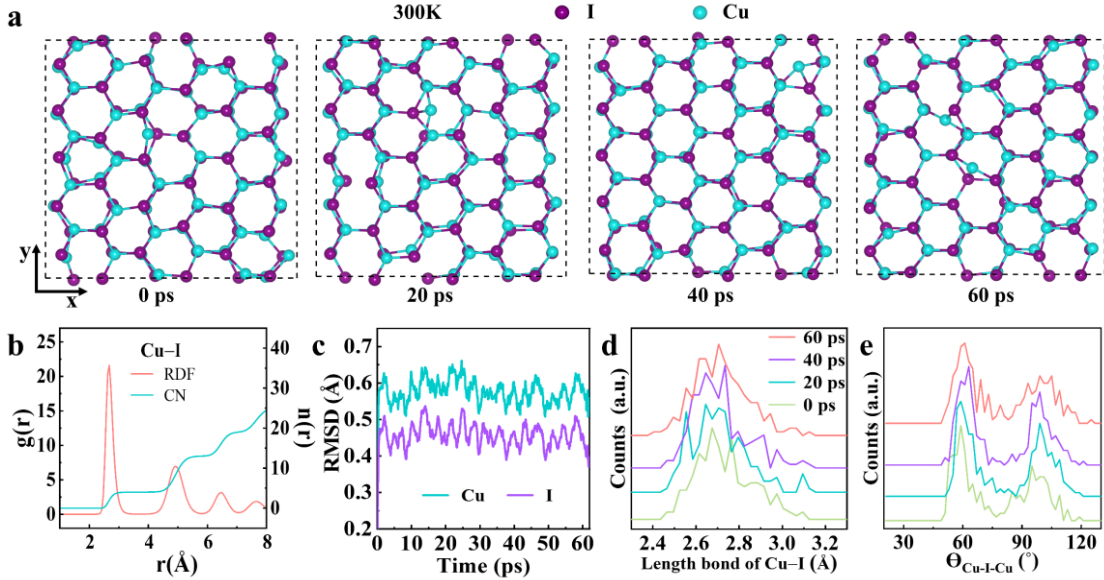

**Figure 3.** AIMD simulations of monolayer β-CuI single crystal at 300 K. (a) Snapshots at 0 ps, 20 ps, 40 ps, and 60 ps. (b) Equilibrated radial distribution function (RDF) and coordination number (CN). (c) Root-mean-square deviations (RMSD) of Cu and I Atoms. (d) Length distribution of Cu–I bond for each snapshot in (a). (e) Statistics of Cu–I–Cu bond angle for each snapshot in (a).

In addition, we investigated the dynamical stability of monolayer β-CuI by computing its phonon spectrum, as shown in **Figure 4**a. According to phonon theory, the presence of negative phonon frequencies indicates structural instability, suggesting a tendency for the structure to transform into a more stable configuration [37]. Clearly, the absence of imaginary frequencies in the first Brillouin zone indicates the inherent dynamical stability of monolayer β-CuI. Additionally, using PBE and HSE06 functionals, we calculated the electronic structure of monolayer β-CuI, revealing band gaps of 2.07 eV and 3.66 eV, respectively. Notably, the valence and conduction bands share the same top and bottom in PBE and HSE06, leading to a direct band-gap, as displayed in Figure 4b. The HSE06 method provides a more accurate description of the electronic structure and band properties of semiconductor materials, particularly in the calculation of band gaps, compared to the traditional PBE method [38-40]. Therefore, monolayer β-CuI can be described as an ultrawide bandgap (UWBG) semiconductor with a direct bandgap of 3.66 eV, surpassing that of γ-CuI (2.95–3.1 eV) [18,41].

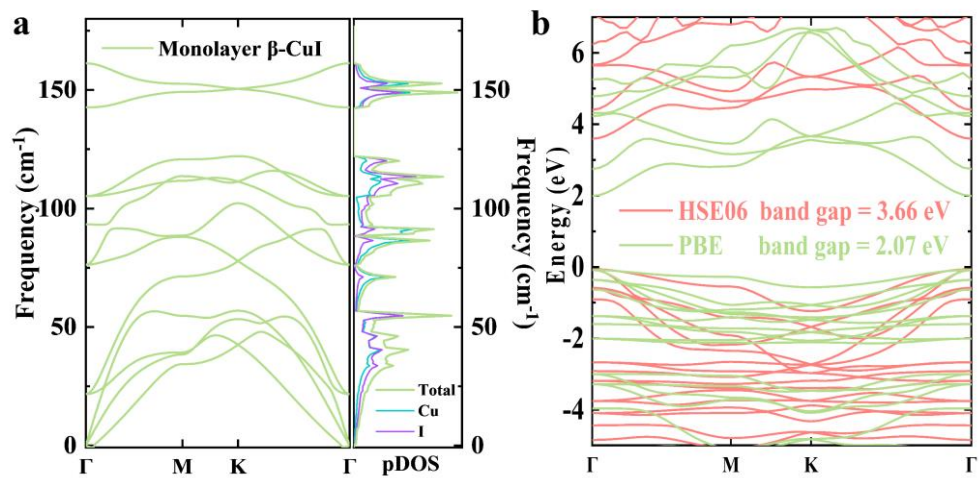

**Figure 4.** Phonon dispersion and band structures of the fully optimized monolayer β-CuI. (a) The phonon dispersions and the partial density of states. (b) Band structure of monolayer β-CuI using PBE and the hybrid functional HSE06.

Interestingly, most 2D ultra-wide bandgap semiconductors developed so far are N-type, but P-type semiconductors are indispensable for the design of many electronic devices [3,42]. However, P-type semiconductors are indispensable for the design of bipolar transistors, inverter circuits, and transparent film transistors [43]. Unfortunately, only a few promising candidates exist as P-type 2D ultra-wide bandgap semiconductors with high mobility [20]. Thus, the exploration of P-type 2D ultra-wide bandgap semiconductors with high mobility is still an urgent problem. β-CuI, as a P-type 2D ultra-wide band-gap semiconductor, has the potential to become a member of the fourth generation of semiconductors [20].

In summary, 2D β-CuI single crystals on the micro-nano scale were found experimentally under ambient conditions, which challenges the previous understanding of β-CuI. By employing suitable processing techniques, we were able to observe trace amounts of 2D β-CuI more effectively within CuI powder using TEM, enabling us to comprehensively characterize and analyze its morphology, structure, and elemental composition. Under continuous electron beam irradiation in high-resolution TEM mode, the stability of 2D β-CuI crystals was tested and found to be relatively stable, allowing for the acquisition of precise atomic-resolved structure images. Our experiment revealed the presence of a moiré superlattice at the point of sample overlap. The theoretical simulation of this superlattice using the Angle theory of phase diagram was consistent with our experimental results, ultimately providing strong evidence for the atomic thickness of 2D β-CuI. The experimental synthesis and meticulous crystal analysis of CuI have unequivocally demonstrated the concurrent coexistence of two distinct structural forms, namely γ-CuI and β-CuI, within the synthesis process. In addition, there were discrepancies in the identification of different phases in synthesized CuI and CuI podwer by XRD and TEM crystal analysis techniques. Specifically, Powder XRD was unable to detect trace amounts of β-CuI in CuI powders, while TEM techniques were shown to be effective under suitable sample preparation conditions.

In addition, AIMD simulations show that monolayer β-CuI is thermally stable at 300 K. Simultaneously, the analysis of the phonon spectra indicates the dynamically stable of monolayer β-CuI. The electronic band structures calculated using the HSE06 functional prove that the monolayer β-CuI has an ultrawide direct bandgap of 3.66 eV. 2D β-CuI has great potential as a P-type 2D ultra-wide band-gap semiconductor and will play a significant role in advancing the development and application of 2D ultra-wide band-gap, P-type transparent semiconductors in academia and industry. The discovery of a β-CuI single crystal at room temperature is unprecedented, which will redefine the phase diagram of CuI at various temperatures. Moreover, this small amount of 2D β-CuI is co-existing with a large amount of γ-CuI in the synthesized CuI under ambient conditions. Due to ineffective characterization, the presence of the β-phase structure has been overlooked, resulting in a lack of understanding of many of its properties, particularly the potential differences and misinterpretations regarding the performance and applications of β-CuI compared to γ-CuI. Our discovery of 2D β-CuI under ambient conditions not only provides a material basis to verify the previously predicted novel properties, but also motivates further experimental studies to investigate the unknown properties of 2D β-CuI and develop potential applications.

**Acknowledgements**

We thank Professor Haiping Fang for his constructive suggestions and helpful discussions, and Yifeng Zheng for his assistance in the phonon spectrum calculation. This work was supported by the National Natural Science Foundation of China (11974366 and 12074341), Startup Fund of Wenzhou Institute, University of Chinese Academy of Sciences (No. WIUCASQD2021014, WIUCASQD2023017), the Fundamental Research Funds for the Central Universities, the Scientific Research and Developed Funds of Ningbo University (No. ZX2022000015).

# Supplementary Information:

### Section 1: Methods

**Section 2: Powder XRD analysis**

**Section 3: The morphology, structure, and elemental characterization of 2D β-CuI crystals through TEM and TEM-EDS analysis**

**Section 4: Crystal diffraction pattern simulation and analysis**

**Section 5: The β and γ phases in the synthesized CuI**

**Section 6: DFT theoretical calculation and AIMD simulations**

**Movie S1**